\documentclass[]{hhr_arxiv}

\title{MINIMAL STABLE VOTING RULES}

\newcommand{\myauthorA}{\textsc{H\'{e}ctor Hermida-Rivera}}%
\newcommand{\myemailA}{\href{mailto:hermida.rivera.hector@gtk.bme.hu}{\texttt{hermida.rivera.hector@gtk.bme.hu}}}%
\newcommand{\myaffiliationA}{\linespread{1.2}\selectfont\emph{Quantitative Social and Management Sciences Research Centre, Faculty of Economic and Social Sciences, Budapest University of Technology and Economics, Muegyetem rkp. 3, Budapest H‑1111, Hungary.}
}%

\begin{document}

\titlesolo

\begin{abstract}
In this paper, I characterize \hyperref[sta]{minimal stable} voting rules and \hyperref[ssta]{minimal self-stable} constitutions (i.e., pairs of voting rules) for societies in which only power matters. To do so, I first let players' preference profiles over voting rules satisfy four natural axioms commonly used in the analysis of power: \hyperref[dom]{non-dominance}, \hyperref[ano]{anonymity}, \hyperref[null]{null player} and \hyperref[sw]{swing player}. I then provide simple notions of \hyperref[sta]{minimal stability} and \hyperref[ssta]{minimal self-stability}, and show that the families of \hyperref[sta]{minimal stable} voting rules and \hyperref[ssta]{minimal self-stable} constitutions are fairly small. Finally, I conclude that political parties have evolved to ensure the \hyperref[ssta]{minimal self-stability} of otherwise not \hyperref[ssta]{minimal self-stable} constitutions.
\end{abstract}

\keywords{voting rules, constitutions, minimal, stability, self-stability, power.}

\jelcodes{D71, D72}

\wordcount{10,015}

% CONTENT

\section{Introduction}\label{sec:intro}

\textsc{Every society} needs a constitution with a voting rule to make collective choices and another to make constitutional amendments.\footnote{Although unusual, both voting rules may be the same.} On the one hand, society's members might wish to set up a constitution that will not be constantly subject to change, as regular transitions are impractical and costly. On the other hand, society's members might not at all be concerned with change, in which case this society's constitution shall naturally evolve until becoming steady. Therefore, there exist both normative and positive reasons to characterize stable voting rules and self-stable constitutions.

For example, in the United Nations General Assembly, most resolutions are passed if and only if they are approved by half of its members, whereas amendments to its Charter are passed if and only if they are approved by two-thirds of its members, including the five UN Security Council permanent members (i.e., France, China, Russia, the UK and the US). In China, ordinary legislation requires a simple majority of the National People's Congress (NPC), yet the reform of its Constitution requires the approval by two of its thirds. In the US, ordinary legislation requires a simple majority of both chambers of its Congress, yet the reform of its Constitution requires the approval by two-thirds of both chambers of its Congress and by three-fourths of its state legislatures.

Formally, a \emph{voting rule} (or \emph{simple game}) is a function that specifies \emph{winning} and \emph{losing coalitions}, where a winning coalition all whose strict sub-coalitions are losing is a \emph{minimal winning coalition} \parencite{vonneumannmorgenstern_44}. The idea is that, like in most legislative bodies around the world, all players face a binary choice between keeping the status quo or replacing it with some alternative, and the alternative gets chosen if and only if all members of some winning coalition vote for it. Since any coalition containing a smaller coalition with the ability to impose the alternative can also impose it, the only formal requirement is that any coalition containing a winning coalition is also winning (i.e., \emph{monotonicity}).\footnote{To prevent contradictions in which two minimal winning coalitions unanimously vote for different alternatives, it is commonly assumed that voting rules are \emph{proper}: namely, that the complement of every winning coalition is losing. However, in binary environments with a single alternative, such contradictions cannot arise: either there is at least one winning coalition all whose members vote for the alternative, or there is none. For a further discussion on this point, see \textcite[pp. 11-13]{felsenthalmachover_98}.}

Clearly, players can be grouped according to their role within any given voting rule: a \emph{swing player} is a player who belongs to some minimal winning coalition, a \emph{null player} is a player who belongs to no minimal winning coalition, a \emph{veto player} is a player who belongs to all minimal winning coalitions, and an \emph{oligarchic player} is a player who belongs to the unique minimal winning coalition whenever there is only one.

In this paper, players' preferences over voting rules satisfy four natural axioms that the weak orderings induced by most power indices also satisfy,\footnote{See \textcite{bertinifreixasgambarellistach_13} for an extensive survey on power indices.} an idea that can be traced back---at least---to \textcite{roth_77a,roth_77b}. According to these four axioms, no voting rule Pareto dominates another (\hyperref[dom]{non-dominance}), only the positions in the voting rules matter (\hyperref[ano]{anonymity}), all players are indifferent among all voting rules in which they are null (\hyperref[null]{null player}), and being swing is strictly better than being null (\hyperref[sw]{swing player}). Together, these axioms are fulfilled by any preference profile represented by any power index satisfying \citeauthor{felsenthalmachover_98}'s (\citeyear{felsenthalmachover_98}, Definition 7.1.1, p. 222) set of necessary postulates for power indices, thus reflecting the preferences of a society whose members simply wish to maximize their own power.

The reader may wonder whether players simply want to maximize their own power. When setting up a constitution, players may not know what alternatives will be voted on, as the decisions society must make are numerous and evolve over time. Similarly, players might correctly anticipate the alternatives but may not know how likely they are to support them, as preferences may evolve over time as well. Confronted with this uncertainty---whether about the alternatives themselves or their likelihood of supporting them---players may simply choose to maximize their own power. It is natural to think heuristically about the family of preference profiles satisfying the four axioms of this paper, as even in complete ignorance, players can always easily compute the power granted to them by each voting rule. Hence, I model the preference behind each player's \emph{veil of ignorance} \parencite{rawls_71} as a desire for power, a modeling choice further supported by the observation that legislators and political parties alike tend to gravitate towards the platforms that maximize their electoral odds.

The first part of this paper analyzes the \hyperref[sta]{minimal stability} requirements of \emph{all voting rules} as defined by \textcite{shapley_62}: namely, all non-null and monotonic simple games.\footnote{All results in this paper hold if the domain is restricted to weighted majority voting rules.} A voting rule is \hyperref[sta]{minimal stable} if and only if there exists no voting rule that Pareto dominates it for any of its minimal winning coalitions. The second part of this paper analyzes the \hyperref[ssta]{minimal self-stability} requirements of constitutions, where a constitution is a pair of voting rules: an \emph{ordinary} one used to decide over regular matters, and an \emph{extraordinary} one used to amend either of its voting rules, \emph{including itself}. A constitution is \hyperref[sta]{minimal self-stable} if and only if there exists no voting rule that Pareto dominates neither its ordinary nor its extraordinary voting rule for any minimal winning coalition of its extraordinary voting rule.\footnote{See \textcite{shapley_62} for an outline of the use of simple games to study voting problems; see \textcite{buchanan_91,voigt_97,voigt_11} for different surveys on constitutional economics.}

This paper contains two central results. \Cref{t.min} states that any voting rule with a unique minimal winning coalition (i.e., \hyperref[oli]{oligarchic}) is \hyperref[sta]{minimal stable}, while any \hyperref[sta]{minimal stable} voting rule has non-empty intersection of minimal winning coalitions (i.e., \hyperref[vet]{veto}). \Cref{t.con} states that a constitution is \hyperref[ssta]{minimal self-stable} if its ordinary swing player set is a subset of its extraordinary oligarchic player set; and it is not \hyperref[ssta]{minimal self-stable} if its extraordinary veto player set is empty, or is neither a subset nor a superset of its ordinary swing player set. So, for example, a constitution whose two voting rules are dictatorships of the same player is \hyperref[ssta]{minimal self-stable}, whereas a constitution requiring a simple majority for ordinary issues and a two-thirds majority for amendments is not \hyperref[ssta]{minimal self-stable}.

The main take-away of this paper is two-fold. On the theoretical side, I show that---quite surprisingly---a collection of four weak axioms severely restricts the sets of \hyperref[sta]{minimal stable} voting rules and \hyperref[ssta]{minimal self-stable} constitutions. On the practical side, I argue that political parties have evolved so as to ensure that most constitutions are \hyperref[ssta]{minimal self-stable}. Hence, this paper provides a novel and simple explanation of why most constitutions around the world undergo relatively few changes. In fact, while constitutional changes do occasionally occur, most such changes pertain to aspects other than the voting rules (e.g., financial stability, fundamental rights, etc.). And historically, whenever the voting rules of a constitution change, they normally do so through extra-constitutional means (e.g., uprisings, wars, etc.). This paper helps us understand why.

Several scholars have studied closely related questions to the one addressed in this paper. However, this paper differs in five notable ways from the works of \textcite{barberajackson_04} and---in particular---\textcite{azrielikim_16}, whose work is the closest to mine. I explain these five differences in detail below. 

First, the characterization of players' preference profiles as a desire for power contrasts with those in which players learn their preferences over voting rules from their preferences over real-world alternatives, as it happens in \textcite{barberajackson_04,azrielikim_16}. In their models of ex-ante self-stability, inspired by \textcite{rae_69}, every player's preference is modeled through some probability distribution indicating the likelihood this player will favor reform. Hence, in their models, players know the alternatives that will be voted on; and although they do not know their preferences over the alternatives, they do know how likely they are to support them. In this paper, players need to know neither the alternatives nor their likelihood of supporting them.

Second, \textcite{barberajackson_04,azrielikim_16} only characterize the stability requirements of (weighted) majority voting rules. In this paper, I follow \citeauthor{shapley_62}'s \citeyearpar{shapley_62} broader conceptualization of a voting rule as a non-null and monotonic simple game, which includes the entire family of weighted majority voting rules. Although many voting rules used in practice are indeed weighted majority voting rules, many important ones are not. Some notable examples are the voting rule of the Council of the European Union\footnote{See \textcite{kirsch_16,freixasmolinero_09} for an analysis of the Council of the European Union.} and the voting rules of numerous bicameral legislatures, as the voting rule that emerges from having two chambers---each with its own (weighted) majority voting rule---is normally not a weighted majority voting rule. 

Third, the definitions of \hyperref[sta]{minimal stability} and \hyperref[ssta]{minimal self-stability} of this paper depart from the self-stability notions of \textcite{barberajackson_04,azrielikim_16}. In these papers, a voting rule is replaced if and only if all players of some of its winning coalitions strictly benefit from the change. In this paper, this is not the case. Therefore, this paper studies stable voting rules and self-stable constitutions in societies that are slightly less averse to change. While there are good reasons to assume some reluctance to change (i.e., transition costs), requiring that every member of some winning coalition strictly benefits from a reform seems sufficient but not necessary to pursue constitutional amendments; for those players who benefit strictly from amending a voting rule may compensate those who are indifferent in other ways (e.g., monetary transfers, policy concessions, etc.).

Fourth, \textcite{barberajackson_04,azrielikim_16} do not consider the possibility that the extraordinary voting rule of a constitution may be used to amend itself. However, most constitutions around the world do indeed allow for this possibility.\footnote{Notable examples of constitutional articles specifying amendment procedures that can be used to amend themselves include Article V of the \textcite{usconstitution_87}, Article 64 of the \textcite{chinaconstitution_82}, Article 108, Chapter XVIII of the \textcite{uncharter_45}, Title X (Articles 166–169) of the \textcite{spainconstitution_78}, Articles 79(1)–(3) of the \textcite{germanybasiclaw_49}, and Article 89 of the \textcite{franceconstitution_58}, to name just a few.} Therefore, I require that the extraordinary voting rule of a constitution be \hyperref[sta]{minimal stable} for a constitution to be \hyperref[ssta]{minimal self-stable}. Dropping this requirement would allow the extraordinary voting rule of a constitution to be amended in such a way that a coalition that desires to change the ordinary voting rule but could not do so before amending the extraordinary voting rule can do so after amending it. Hence, it seems reasonable to require that the extraordinary voting rule of a constitution will not be replaced for such a constitution to be \hyperref[ssta]{minimal self-stable}.

Fifth, the results of this paper contrast with those obtained by \textcite{barberajackson_04,azrielikim_16}. In their analysis of ex-ante self-stable anonymous constitutions, \textcite{barberajackson_04} show that there exist societies for which constitutions with a majoritarian extraordinary voting rule are self-stable, as well as societies with no self-stable constitutions; in this paper, neither conclusion holds. However, they argue that a constitution with a majoritarian ordinary voting rule and the unanimous extraordinary voting rule is self-stable, a conclusion this paper agrees with. In their extension of \citeauthor{barberajackson_04}'s \citeyearpar{barberajackson_04} model to weighted majority voting rules, \textcite{azrielikim_16} show that almost every stable voting rule must be a \hyperref[vet]{veto} voting rule, a similar---but not identical---conclusion to the one this paper obtains (see \Cref{ex1,ex2}). Furthermore, \textcite{azrielikim_16} derive some necessary and sufficient conditions for a constitution to be self-stable. While the necessary conditions I identify in this paper agree with theirs, I also obtain different sufficient conditions for a constitution to be \hyperref[ssta]{minimal self-stable} (see \Cref{ex3,ex4}). 

Finally, some authors have taken further departures from \citeauthor{barberajackson_04}'s \citeyearpar{barberajackson_04} seminal model. For example, \textcite{jeongkim_23} introduce the notions of interim and ex-post self-stability, by which players privately (and publicly) know their preferences over real-world outcomes. They show that interim self-stable constitutions are quite robust, and that interim and ex-post self-stable anonymous constitutions coincide (but differ from ex-ante self-stable ones). In a related paper, \textcite{jeongkim_24} close the gap between ex-post and ex-ante self-stability. Further, \textcite{kulttimiettinen_07} consider an infinite player set, and characterize the \textcite{vonneumannmorgenstern_44} stable set of majority voting rules, which always exists, is unique and contains the set of self-stable voting rules as defined by \textcite{barberajackson_04}. \textcite{coelho_05} introduces a probabilistic setting and shows that the set of voting rules satisfying the maximin criterion is often different from the majority voting rule. \textcite{kulttimiettinen_09} consider constitutions with an arbitrary number of voting rules, each used to amend the previous one; and show that self-stable constitutions generally contain the same voting rule from the second onward. Lastly, \textcite{lagunoff_09} introduces a dynamic model in which voting rules are self-selective if and only if they satisfy a condition of dynamic consistency. 

This paper is organized as follows: \Cref{sec:env} defines the environment and players' preferences, \Cref{sec:sta} characterizes \hyperref[sta]{minimal stable} voting rules, \Cref{sec:self} characterizes \hyperref[ssta]{minimal self-stable} constitutions, and \Cref{sec:con} concludes with some real-world examples.

\section{Environment}\label{sec:env}

The \emph{environment} is a $3$-tuple $(N,\mathcal{V},\mathcal{R})$, where $N=\{1,\dots,n\}$ is a \emph{finite player set};
\begin{gather}
    \mathcal{V}=\big\{\nu:2^N\to\{0,1\}\mid(\nu(\emptyset)=0)\wedge(\nu(N)=1)\wedge[(T\subseteq T')\Rightarrow(\nu(T)\leqslant\nu(T'))]\big\}
\end{gather}
is the \emph{set of all voting rules} (i.e., non-null and monotonic simple games), each of which is a function that assigns either $0$ or $1$ to every coalition of players; and
\begin{gather}
    \mathcal{R}=\{R\mid R=(R_i)_{i\in N}: R_i\text{ is player $i$'s \emph{weak ordering} over }\mathcal{V}\}
\end{gather}
is the \emph{set of all preference profiles}\footnote{A \emph{weak ordering} is a \emph{complete}, \emph{transitive} and \emph{reflexive} binary relation.} over $\mathcal{V}$, where $\nu R_i\nu'$ if and only if player $i$ \emph{weakly prefers} $\nu$ over $\nu'$, $\nu P_i\nu'$ if and only if player $i$ \emph{strictly prefers} $\nu$ over $\nu'$, and $\nu I_i\nu'$ if and only if player $i$ \emph{is indifferent} between $\nu$ and $\nu'$.

\subsection{Voting rules}\label{sub:rules}

\Cref{sub:rules} introduces some notation and concepts related to voting rules. Given any voting rule $\nu\in \mathcal{V}$, let $\mathcal{W}(\nu)=\{T\subseteq N\mid\nu(T)=1\}$ be its \emph{set of winning coalitions} (i.e., the set of coalitions whose worth is one), and let $\mathcal{M}(\nu)=\{T\in\mathcal{W}(\nu)\mid(\forall L\subsetneq T)(\nu(L)=0)\}$ be its \emph{set of minimal winning coalitions} (i.e., the set of winning coalitions all whose strict sub-coalitions have zero worth).

Further, let $S(\nu)=\bigcup_{T\in\mathcal{M}(\nu)}T$ be its \emph{swing player set} (i.e., the set of players who belong to some minimal winning coalition),\footnote{\emph{Swing players} are also called \emph{pivot}, \emph{pivotal}, \emph{crucial} and \emph{critical players}.} let $N(\nu)=\{i\in N\mid i\in N\backslash S(\nu)\}$ be its \emph{null player set} (i.e., the set of players who belong to no minimal winning coalition), let $V(\nu)=\bigcap_{T\in\mathcal{M}(\nu)}T$ be its \emph{veto player set} (i.e., the set of players who belong to all minimal winning coalitions),\footnote{\emph{Veto players} are also called \emph{blocking players}.} and let 
\begin{gather}
O(\nu)=
\begin{cases}
    S(\nu)&\text{if }|\mathcal{M}(\nu)|=1\\
    \emptyset&\text{else}
\end{cases}
\end{gather}
be its \emph{oligarchic player set} (i.e., the set of players who belong to the unique minimal winning coalition whenever there is only one).

\begin{definition}[Veto]\label{vet}
    A voting rule $\nu$ is \emph{veto} if and only if it has veto players. Formally, if and only if
\begin{gather}
    V(\nu)\neq\emptyset
\end{gather}
\end{definition}

Let $\mathcal{V}_v=\{\nu\in \mathcal{V}\mid\nu\text{ is \hyperref[vet]{veto}}\}$.

\begin{definition}[Oligarchy]\label{oli}
    A voting rule $\nu$ is \emph{oligarchic} if and only if it has oligarchic players. Formally, if and only if
\begin{gather}
    O(\nu)\neq\emptyset
\end{gather}
\end{definition}

Let $\mathcal{V}_o=\{\nu\in \mathcal{V}\mid\nu\text{ is \hyperref[oli]{oligarchic}}\}$.\footnote{\emph{\hyperref[oli]{Oligarchic} voting rules} are also called \emph{unanimity games}.}

\subsection{Preference profile}\label{sub:ut}

\Cref{sub:ut} introduces four natural axioms that capture the preferences of a society whose members are exclusively concerned with maximizing their own power. Then, it discusses how these axioms relate to the existing literature on power measurement.

Let $\nu\succcurlyeq_T\nu'$ if and only if $\nu$ \emph{Pareto dominates $\nu'$ for coalition $T$ at preference profile $R$}: formally,
\begin{gather}
    (\nu\succcurlyeq_T\nu')\iff[((\forall i\in T)(\nu R_i\nu'))\wedge((\exists i\in T)(\nu P_i\nu'))]
\end{gather}

\begin{axiom}[Non-dominance]\label{dom}
A preference profile $R$ satisfies the \emph{non-dominance} axiom if and only if no voting rule Pareto dominates another. Formally, if and only if
\begin{gather}
    (\nexists\nu,\nu'\in \mathcal{V})(\nu\succcurlyeq_N\nu')
\end{gather}
\end{axiom}

Let $\Pi=\{\pi:N\to N\mid\pi\text{ is bijective}\}$ be the \emph{set of all permutations of the player set}. Then, given any voting rule $\nu\in \mathcal{V}$ and any permutation $\pi\in\Pi$, let $\pi\nu\in \mathcal{V}$ be the \emph{isomorphic voting rule} defined by $\pi\nu(T)=\nu(\pi(T))$ for all coalitions $T\subseteq N$ (i.e., the voting rule that just swaps players' names according to the permutation $\pi$).

\begin{axiom}[Anonymity]\label{ano}
    A preference profile $R$ satisfies the \emph{anonymity} axiom if and only if it is independent of players' names. Formally, if and only if
\begin{equation}
\begin{gathered}
    (\forall\nu,\nu'\in \mathcal{V})(\forall\pi\in\Pi)(\forall i\in N)[(\nu R_i\nu')\iff(\pi\nu R_{\pi(i)}\pi\nu')]
\end{gathered}
\end{equation}
\end{axiom}

\begin{axiom}[Null player]\label{null}
    A preference profile $R$ satisfies the \emph{null player} axiom if and only if all players are indifferent among all voting rules in which they are null. Formally, if and only if
\begin{gather}
    (\forall\nu,\nu'\in \mathcal{V})(\forall i\in N(\nu)\cap N(\nu'))(\nu I_i\nu')
\end{gather}
\end{axiom}

\begin{axiom}[Swing player]\label{sw}
    A preference profile $R$ satisfies the \emph{swing player} axiom if and only if all players strictly prefer being swing over being null. Formally, if and only if
\begin{gather}
    (\forall\nu,\nu'\in \mathcal{V})(\forall i\in S(\nu)\cap N(\nu'))(\nu P_i\nu')
\end{gather}
\end{axiom}

Let $\mathcal{R}^*=\{R\in\mathcal{R}\mid R\text{ satisfies \cref{ano,sw,null,dom}}\}$.

A \emph{power measure} is a function $\mu:\mathcal{V}\to\mathbb{R}^N_+$ that quantifies each player's power in every voting rule; if in addition it is normalized to $1$, it is a \emph{power index} $\rho:\mathcal{V}\to[0,1]^N$. Given that every power index represents a preference profile $R\in\mathcal{R}$ satisfying $\nu R_i\nu'$ if and only if $\rho_i(\nu)\geqslant\rho_i(\nu')$, \Cref{indices} summarizes which power indices of those surveyed by \textcite{bertinifreixasgambarellistach_13} induce a preference profile satisfying \Cref{ano,dom,sw,null}. Notably, the indices in \Cref{indices} marked with an asterisk are (non-normalized) measures. For such measures, \Cref{indices} considers their normalization given by $\rho_i(\nu)=\mu_i(\nu)/\sum_{j\in N}\mu_j(\nu)$ for all players $i\in N$ and all voting rules $\nu\in \mathcal{V}$. Since the \textcite{banzhaf_64} power index is the normalized \textcite{penrose_46} power measure, \Cref{indices} excludes the latter; however, it includes \citeauthor{schmeidler_69}'s \citeyearpar{schmeidler_69} Nucleolus, which was characterized by \textcite{montero_13} as a power index. 

\begin{table}
\caption{Axioms of key preference profiles}
\begin{threeparttable}
\centering
\renewcommand{\arraystretch}{1}
\begin{tabularx}{\textwidth}{rXcXcXcXc}\hline\hline
 \textsc{Power index} & & \textsc{A1} & & \textsc{A2} & & \textsc{A3} & & \textsc{A4} \\\hline
 \textcite{shapleyshubik_54} & & yes  & & yes  & & yes  & & yes  \\
 \textcite{banzhaf_64} & & yes  & & yes  & & yes  & & yes  \\
 Nucleolus \parencite{schmeidler_69} & & yes  & & yes  & & yes  & & \emph{no}\\
 \textcite{rae_69} * & & yes & & yes  & & \emph{yes}\textsuperscript{\textdagger}  & & yes  \\
 \textcite{coleman_71} to initiate * & & yes  & & yes  & & yes  & & yes  \\
 \textcite{coleman_71} to prevent * & & yes  & & yes  & & yes  & & yes  \\
 \textcite{deeganpackel_78} & & yes  & & yes  & & yes  & & yes  \\
 \textcite{johnston_78} & & yes  & & yes  & & yes  & & yes  \\
 \textcite{nevisonzichtshoepke_78}\textsuperscript{a} * & & yes  & & yes  & & \emph{yes}\textsuperscript{\textdagger}  & & yes \\
 Regular semivalues \parencite{weber_79} * & & yes  & & yes  & & yes  & & yes \\
 Non-regular semivalues \parencite{weber_79} * & & yes  & & yes  & & yes  & & \emph{no} \\
 $\tau$-value \parencite{tijs_81} & & yes  & & yes  & & yes  & & \emph{no} \\
 Public good index \parencite{holler_82} & & yes  & & yes  & & yes  & & yes  \\
 \textcite{bertinigambarellistach_08}\textsuperscript{b} & & yes  & & yes  & & \emph{no} & & yes  \\
 \textcite{alonsomeijidefreixas_10}\textsuperscript{c} & & yes  & & yes  & & yes  & & yes  \\\hline\hline
\end{tabularx}
\begin{tablenotes}[flushleft]
\linespread{1.2}\footnotesize
\item[]\textsuperscript{\textdagger}These two measures $\mu:\mathcal{V}\to\mathbb{R}_+^N$ satisfy $\mu_i(\nu)=(1/2)$ for all null players $i\in N(\nu)$. Therefore, the preference profiles represented by their normalizations do not satisfy the \hyperref[null]{null player} axiom. However, the measures $\mu':\mathcal{V}\to\mathbb{R}_+^N$ given by $\mu_i'(\nu)=0$ for all null players $i\in N(\nu)$ and $\mu_i'(\nu)=\mu_i(\nu)$ for all swing players $i\in S(\nu)$ represent the same preference profiles as the original measures, while the preference profiles represented by their normalizations satisfy the \hyperref[null]{null player} axiom.\vspace{2pt}
\item[]\textsuperscript{a}Also known as the \emph{inclusiveness index}.\vspace{2pt}
\item[]\textsuperscript{b}Also known as the \emph{public help index}.\vspace{2pt}
\item[]\textsuperscript{c}Also known as the \emph{shift power index}.
\end{tablenotes}
\end{threeparttable}
\label{indices}
\end{table}

In an exhaustive book on power measurement, \textcite[Definition 7.1.1, p. 222]{felsenthalmachover_98} laid out a short and widely accepted collection of necessary postulates for any power index to qualify as such. It turns out that \Cref{dom,null,sw,ano} are satisfied by every preference profile represented by any power index satisfying their postulates. Clearly, the \hyperref[dom]{non-dominance} axiom is satisfied by all preference profiles represented by some power index, as these are non-negative and normalized functions. The \hyperref[ano]{anonymity} axiom is satisfied by every preference profile represented by any power index satisfying \citeauthor{felsenthalmachover_98}'s (\citeyear{felsenthalmachover_98}, Postulate 1, p. 222) \emph{iso-invariance}: namely, all of them. Further, any preference profile represented by some power index satisfying \citeauthor{felsenthalmachover_98}'s (\citeyear{felsenthalmachover_98}, Postulate 3, p. 222) \emph{vanishing just for dummies} satisfies the \hyperref[null]{null player} and \hyperref[sw]{swing player} axioms. It turns out that both axioms are satisfied by the preference profiles represented by most---but not all---power indices. Finally, \textcite[Postulate 2, p. 222]{felsenthalmachover_98} introduce one additional postulate without any role in this paper (i.e., \emph{ignoring dummies}).

\section{Minimal stable voting rules}\label{sec:sta}

The first part of \Cref{sec:sta} introduces a natural notion of \hyperref[sta]{minimal stability} and characterizes \hyperref[sta]{minimal stable} voting rules when players' preference profile satisfies \Cref{dom,ano,null,sw}. The second part of \Cref{sec:sta} introduces two alternative notions of stability and finds families of voting rules that satisfy (and do not satisfy) these definitions when players' preference profile satisfies \Cref{dom,ano,null,sw}.

\subsection{Main results}\label{sub:res}

\begin{definition}[Minimal stability]\label{sta}
    A voting rule $\nu$ is \emph{minimal stable in $R$} if and only if there exists no voting rule that Pareto dominates it for any of its minimal winning coalitions. Formally, if and only if
\begin{gather}\tag{ms}\label{eq:sta}
    (\forall\nu'\in \mathcal{V})[(\nexists T\in\mathcal{M}(\nu))(\nu' \succcurlyeq_T\nu)]
\end{gather}
\end{definition}

Let $\mathcal{V}_s(R)=\{\nu\in \mathcal{V}\mid\nu\text{ is \hyperref[sta]{minimal stable in $R$}}\}$.

To understand the idea behind \hyperref[sta]{minimal stability}, suppose there is some incumbent voting rule and that players vote either in favor or against some alternative voting rule. Then, a player votes for the alternative if he belongs to some minimal winning coalition of the incumbent voting rule for which the alternative constitutes a Pareto improvement. Finally, the alternative voting rule is adopted if and only if there exists a minimal winning coalition of the incumbent voting rule all whose members vote for it. Therefore, this novel stability notion is loosely inspired by \citeauthor{riker_62}'s \citeyearpar{riker_62} theory of coalition formation, according to which only minimal winning coalitions will rule. When compared with other stability notions found in the literature, the main novelty is the use of standard---rather than strong---Pareto domination. 

\begin{theorem}\label{t.min}
    Let players' preference profile satisfy the \hyperref[dom]{non-dominance}, \hyperref[ano]{anonymity}, \hyperref[null]{null player} and \hyperref[sw]{swing player} axioms. Then, the set of \hyperref[sta]{minimal stable} voting rules contains the set of \hyperref[oli]{oligarchic} voting rules and is contained in the set of \hyperref[vet]{veto} voting rules. Formally,
\begin{gather}
    (\forall R\in \mathcal{R}^*)[\mathcal{V}_o\subseteq \mathcal{V}_s(R)\subseteq \mathcal{V}_v]
\end{gather}
\end{theorem}

\begin{proof}
    Consider any preference profile $R\in\mathcal{R}^*$. Then, there are two statements to show:
\begin{enumerate}
    \item $\mathcal{V}_o\subseteq \mathcal{V}_s(R)$,
    \item $\mathcal{V}_s(R)\subseteq \mathcal{V}_v$.
\end{enumerate}

\begin{statement}
    $\mathcal{V}_o\subseteq \mathcal{V}_s(R)$.   
\end{statement}

    The proof is direct. Consider any \hyperref[oli]{oligarchic} voting rule $\nu\in \mathcal{V}_o$. By the \hyperref[sw]{swing player} axiom, there exists some voting rule $\nu'\in \mathcal{V}$ satisfying $\nu\neg I_i\nu'$ for some player $i\in N$. Fix any such voting rule $\nu'\in \mathcal{V}$. By the \hyperref[dom]{non-dominance} axiom, there exist two players $i,j\in N$ such that $\nu P_i\nu'$ and $\nu' P_j\nu$. By the \hyperref[null]{null player} axiom, $\nu'I_i\nu$ for all players $i\in N(\nu')\backslash O(\nu)$. By the \hyperref[sw]{swing player} axiom, $\nu' P_i\nu$ for all players $i\in S(\nu')\backslash O(\nu)$. Since $N(\nu')\cup S(\nu')=N$, it follows that $\nu'R_i\nu$ for all non-oligarchic players $i\in N\backslash O(\nu)$. Hence, $\nu P_i\nu'$ for some oligarchic player $i\in O(\nu)$. Since $\mathcal{M}(\nu)=\{O(\nu)\}$, there exists no minimal winning coalition $T\in\mathcal{M}(\nu)$ satisfying $\nu'\succcurlyeq_T\nu$. Therefore, \cref{eq:sta} is satisfied. Thus, $\nu\in \mathcal{V}_s(R)$. And finally, $\mathcal{V}_o\subseteq \mathcal{V}_s(R)$.
    
\begin{statement}
    $\mathcal{V}_s(R)\subseteq \mathcal{V}_v$.   
\end{statement}
    
    Given any voting rule $\nu\in \mathcal{V}$, let $\tilde{\Pi}(\nu)=\{\pi\in\Pi\mid \pi(S(\nu))=S(\nu)\}$ be the \emph{set of all permutations of the player set mapping the swing player set to itself}. Further, define the following two exhaustive groups of non-\hyperref[vet]{veto} voting rules: 
\begin{gather}
    \mathcal{V}_1=\{\nu\in\mathcal{V}\backslash\mathcal{V}_v\mid(\forall i\in S(\nu))(\forall\pi\in\tilde{\Pi}(\nu))(\nu I_i\pi\nu)\}\\
    \mathcal{V}_2=\{\nu\in\mathcal{V}\backslash\mathcal{V}_v\mid(\exists i\in S(\nu))(\exists\pi\in\tilde{\Pi}(\nu))(\nu\neg I_i\pi\nu)\}
\end{gather}
    Since $\mathcal{V}\backslash \mathcal{V}_v=\mathcal{V}_1\cup \mathcal{V}_2$, there are two claims to show:
\begin{enumerate}[label=2.\arabic*.]  
    \item $\mathcal{V}_1\subseteq \mathcal{V}\backslash \mathcal{V}_s(R)$,
    \item $\mathcal{V}_2\subseteq \mathcal{V}\backslash \mathcal{V}_s(R)$.
\end{enumerate}

\begin{claim}\label{t.min:c21}
    $\mathcal{V}_1\subseteq \mathcal{V}\backslash \mathcal{V}_s(R)$.
\end{claim}

    The proof is direct. Consider any voting rule $\nu\in \mathcal{V}_1$ and any minimal winning coalition $T\in\mathcal{M}(\nu)$. Then, consider the unique voting rule $\nu'\in \mathcal{V}$ satisfying $\mathcal{M}(\nu')=\{T\}$. Since $N(\nu)\subsetneq N(\nu')=T^c$, the \hyperref[null]{null player} axiom implies $\nu'I_i\nu$ for all null players $i\in N(\nu)$, whereas the \hyperref[sw]{swing player} axiom implies $\nu P_i\nu'$ for all players $i\in S(\nu)\cap N(\nu')$. Thus, $\nu\succcurlyeq_{T^c}\nu'$. Then, by the \hyperref[dom]{non-dominance} axiom, $\nu' P_i\nu$ for some player $i\in T$. Since $\pi\nu I_i\nu$ for all swing players $i\in S(\nu)$ and all permutations $\pi\in\tilde{\Pi}(\nu)$, the \hyperref[ano]{anonymity} axiom implies $\nu' P_i\nu$ for all players $i\in T$. Hence, $\nu'\succcurlyeq_T\nu$. Since $T\in\mathcal{M}(\nu)$, it follows that \cref{eq:sta} is not satisfied. Thus, $\nu\in\mathcal{V}\backslash\mathcal{V}_s(R)$. Therefore, $\mathcal{V}_1\subseteq \mathcal{V}\backslash \mathcal{V}_s(R)$.

\begin{claim}\label{t.min:c22}
    $\mathcal{V}_2\subseteq \mathcal{V}\backslash \mathcal{V}_s(R)$.
\end{claim}

    The proof is direct. Consider any voting rule $\nu\in \mathcal{V}_2$. By the \hyperref[null]{null player} axiom, $\nu I_i\pi\nu$ for all null players $i\in N(\nu)$ and all permutations $\pi\in\tilde{\Pi}(\nu)$. By the \hyperref[dom]{non-dominance} axiom, there exist two swing players $i,j\in S(\nu)$ such that $\pi\nu P_i\nu$ and $\nu P_j\pi\nu$ for some permutation $\pi\in\tilde{\Pi}(\nu)$. Since $V(\nu)=\emptyset$, the \hyperref[ano]{anonymity} axiom implies that there exists some minimal winning coalition $T\in\mathcal{M}(\nu)$ and one such pair of swing players $i,j\in S(\nu)$ such that player $i\in T$ and player $j\in N\backslash T$. Fix any such minimal winning coalition $T\in\mathcal{M}(\nu)$ and any such two swing players $i,j\in S(\nu)$. Consider the permutation $\pi\in\Pi$ satisfying $\pi(i)=j$, $\pi(j)=i$ and $\pi(k)=k$ for all players $k\in N\backslash\{i,j\}$. By the \hyperref[ano]{anonymity} axiom, $\pi\nu P_i\nu$, $\nu P_j\pi\nu$ and $\pi\nu I_k\nu$ for all players $k\in N\backslash\{i,j\}$. Since player $i\in T$ and player $j\in N\backslash T$, it follows that $\pi\nu\succcurlyeq_T\nu$. Since $T\in\mathcal{M}(\nu)$, \cref{eq:sta} is not satisfied. Hence, $\nu\in\mathcal{V}\backslash \mathcal{V}_s(R)$. Therefore, $\mathcal{V}_2\subseteq \mathcal{V}\backslash \mathcal{V}_s(R)$.
\end{proof}

The reader may wonder whether the upper and lower bounds identified in \Cref{t.min} are sharp. In \Cref{clcomp1,clcomp2}, I show they are. Therefore, a complete characterization of \hyperref[sta]{minimal stable} voting rules cannot be obtained without additional axioms.

\begin{corollary}\label{clcomp1}
    There exists a preference profile satisfying the \hyperref[dom]{non-dominance}, \hyperref[ano]{anonymity}, \hyperref[null]{null player} and \hyperref[sw]{swing player} axioms for which the set of \hyperref[sta]{minimal stable} voting rules coincides with the set of  \hyperref[oli]{oligarchic} voting rules. Formally,
\begin{gather}\label{coincide1}
    (\exists R\in\mathcal{R}^*)(\mathcal{V}_s(R)=\mathcal{V}_o)
\end{gather}
\end{corollary}

\begin{cproof}
    The proof is direct. Consider the preference profile $R\in\mathcal{R}$ represented by the power index $\rho:\mathcal{V}\to[0,1]^N$ satisfying, for all voting rules $\nu\in \mathcal{V}$, $\rho_i(\nu)=1/|S(\nu)|$ if player $i\in S(\nu)$, and $\rho_i(\nu)=0$ if player $i\in N(\nu)$. Clearly, $R\in\mathcal{R}^*$. By \Cref{t.min}, $\mathcal{V}_o\subseteq \mathcal{V}_s(R)\subseteq \mathcal{V}_v$. Hence, consider any non-\hyperref[oli]{oligarchic} \hyperref[vet]{veto} voting rule $\nu\in \mathcal{V}_v\backslash \mathcal{V}_o$ and any voting rule $\nu'\in \mathcal{V}$ satisfying $N(\nu')=N(\nu)\cup\{i\}$ for some non-veto swing player $i\in S(\nu)\backslash V(\nu)$. Then, $\nu P_i\nu'$ and $\nu'P_j\nu$ for all swing players $j\in S(\nu)\backslash\{i\}$. Since $i\in N\backslash V(\nu)$, fix any minimal winning coalition $T\in\mathcal{M}(\nu)$ for which player $i\in N\backslash T$. Then, $\nu'\succcurlyeq_T\nu$. Hence, \cref{eq:sta} is not satisfied. Then, $\nu\in \mathcal{V}\backslash \mathcal{V}_s(R)$. Thus, $\mathcal{V}_s(R)=\mathcal{V}_o$.
\end{cproof}

\begin{corollary}\label{clcomp2}
    There exists a preference profile satisfying the \hyperref[dom]{non-dominance}, \hyperref[ano]{anonymity}, \hyperref[null]{null player} and \hyperref[sw]{swing player} axioms for which the set of \hyperref[sta]{minimal stable} voting rules coincides with the set of \hyperref[vet]{veto} voting rules. Formally,
\begin{gather}\label{coincide2}
    (\exists R\in \mathcal{R}^*)(\mathcal{V}_s(R)=\mathcal{V}_v)
\end{gather}
\end{corollary}

\begin{cproof}
    The proof is by contradiction. Let $\varepsilon>0$ be arbitrarily small. Now, consider the preference profile $R\in\mathcal{R}$ represented by the power index $\rho:\mathcal{V}\to[0,1]^N$ satisfying, for all non-\hyperref[vet]{veto} voting rules $\nu\in\mathcal{V}\backslash \mathcal{V}_v$, $\rho_i(\nu)=1/|S(\nu)|$ if player $i\in S(\nu)$, and $\rho_i(\nu)=0$ else; as well as, for all \hyperref[vet]{veto} voting rules $\nu\in \mathcal{V}_v$,
\begin{gather}
    \rho_i(\nu)=
\begin{dcases}
    \frac{\varepsilon}{|S(\nu)|^{|S(\nu)|}}&\text{if player }i\in S(\nu)\backslash V(\nu)\\
    \frac{1}{|V(\nu)|}\left(1-\sum_{j\in S(\nu)\backslash V(\nu)}\rho_j(\nu)\right)&\text{if player }i\in V(\nu)\\
    0&\text{else}
\end{dcases}
\end{gather}
    Since the power index $\rho:\mathcal{V}\to[0,1]^N$ is independent of players' names (i.e., it only varies with the players' role in the voting rule), the preference profile $R$ satisfies the \hyperref[ano]{anonymity} axiom. Since $V(\nu)\subseteq S(\nu)$ and $N\backslash S(\nu)=N(\nu)$, it follows that $\rho_i(\nu)=0$ for all null players $i\in N(\nu)$ and all voting rules $\nu\in \mathcal{V}$. Hence, the preference profile $R$ satisfies the \hyperref[null]{null player} axiom. For $\varepsilon>0$ sufficiently small, $\rho_i(\nu)>0$ for all swing players $i\in S(\nu)$ and all voting rules $\nu\in \mathcal{V}$. Hence, the preference profile $R$ satisfies the \hyperref[sw]{swing player} axiom if $\varepsilon>0$ is sufficiently small. Finally, for $\varepsilon>0$ sufficiently small, we have $\sum_{i\in N}\rho_i(\nu)=1$ for all voting rules $\nu\in\mathcal{V}$. Thus, since $\rho_i(\nu)\geqslant0$ for all players $i\in N$ and all voting rules $\nu\in\mathcal{V}$ if $\varepsilon>0$ is sufficiently small, it follows that the preference profile $R$ satisfies the \hyperref[dom]{non-dominance} axiom if $\varepsilon>0$ is sufficiently small. Therefore, $R\in\mathcal{R}^*$ if $\varepsilon>0$ is sufficiently small. 
    
    By \Cref{t.min}, $\mathcal{V}_o\subseteq \mathcal{V}_s(R)\subseteq \mathcal{V}_v$. Hence, consider any non-\hyperref[oli]{oligarchic} \hyperref[vet]{veto} voting rule $\nu\in \mathcal{V}_v\backslash \mathcal{V}_o$ and suppose that some voting rule $\nu'\in \mathcal{V}$ satisfies $\nu'\succcurlyeq_T\nu$ for some minimal winning coalition $T\in\mathcal{M}(\nu)$. For $\varepsilon>0$ sufficiently small, it follows that $\nu'\in\mathcal{V}_v\backslash \mathcal{V}_o$ and $V(\nu')=V(\nu)$. Thus, $|S(\nu')|\neq|S(\nu)|$. Suppose that $|S(\nu')|>|S(\nu)|$. Then, $\rho_i(\nu')<\rho_i(\nu)$ for all non-veto swing players $i\in S(\nu)\backslash V(\nu)$. Since $(S(\nu)\backslash V(\nu))\cap T\neq\emptyset$ for all minimal winning coalitions $T\in\mathcal{M}(\nu)$, there exists no minimal winning coalition $T\in\mathcal{M}(\nu)$ such that $\nu'\succcurlyeq_T\nu$. Hence, suppose that $|S(\nu')|<|S(\nu)|$. Then, it suffices to verify that
\begin{gather}\label{key}
    1-\sum_{i\in S(\nu)\backslash V(\nu)}\rho_i(\nu)>1-\sum_{i\in S(\nu')\backslash V(\nu)}\rho_i(\nu')
\end{gather}
    Now, \cref{key} is equivalent to \cref{new}:
\begin{align}\label{new}
    |S(\nu)\backslash V(\nu)||S(\nu)|^{-|S(\nu)|}<|S(\nu')\backslash V(\nu)||S(\nu')|^{-|S(\nu')|}
\end{align}
    To see that \cref{new} holds, re-arrange it to
\begin{gather}
    |S(\nu)\backslash V(\nu)||S(\nu')|^{|S(\nu')|}<|S(\nu')\backslash V(\nu)||S(\nu)|^{|S(\nu)|}
\end{gather}
    Since $|V(\nu)|\leqslant|S(\nu')|-2$, a simpler but stricter condition is
\begin{gather}
    |S(\nu)\backslash V(\nu)||S(\nu')|^{|S(\nu')|}<2 |S(\nu)|^{|S(\nu)|}
\end{gather}
    Since $|V(\nu)|\geqslant1$, a simpler but stricter condition is
\begin{gather}\label{last0}
    |S(\nu)||S(\nu')|^{|S(\nu')|}<2 |S(\nu)|^{|S(\nu)|}
\end{gather}
    Now, \cref{last0} is equivalent to \cref{last}:
\begin{gather}\label{last}
    |S(\nu')|^{|S(\nu')|}<2|S(\nu)|^{|S(\nu)|-1}
\end{gather}
    Since $|S(\nu)|-1\geqslant|S(\nu')|>1$, \cref{last} holds. Therefore, \cref{key} holds. Thus, $\rho_i(\nu')<\rho_i(\nu)$ for all veto players $i\in V(\nu)$. Since $V(\nu)\subsetneq T$ for all minimal winning coalitions $T\in\mathcal{M}(\nu)$, there exists no minimal winning coalition $T\in\mathcal{M}(\nu)$ satisfying $\nu'\succcurlyeq_T\nu$. Then, \cref{eq:sta} is satisfied. Hence, $\nu\in \mathcal{V}_s(R)$. Thus, $\mathcal{V}_s(R)=\mathcal{V}_v$.
\end{cproof}

Finally, in \Cref{ex1,ex2}, I illustrate how the results in \Cref{t.min} and \Cref{clcomp1,clcomp2} compare with those of \textcite[Theorem 1, p. 379]{azrielikim_16}.

\begin{exampleb}\label{ex1}
    Consider a society with three players $N=\{1,2,3\}$ and the \hyperref[oli]{oligarchic} voting rule $\nu\in\mathcal{V}_o$ given by the unique minimal winning coalition $\mathcal{M}(\nu)=\{\{1,2\}\}$. By \textcite[Theorem 1i(A) \& 1ii, p. 379]{azrielikim_16}, the voting rule $\nu$ is self-stable if and only if players' utility function is parametrized by $r\in(0,1]$. However, since $\nu\in\mathcal{V}_o$, it follows from \Cref{t.min} that the voting rule $\nu$ is \hyperref[sta]{minimal stable} for all preference profiles satisfying \cref{ano,null,sw,dom}.
\end{exampleb}

\begin{exampleb}\label{ex2}
    Consider a society with three players $N=\{1,2,3\}$ and the non-\hyperref[vet]{veto} voting rule $\nu\in\mathcal{V}\backslash\mathcal{V}_v$ given by the minimal winning coalitions $\mathcal{M}(\nu)=\{\{1,2\},\{1,3\},\{2,3\}\}$. By \textcite[Theorem 1ii, p. 379]{azrielikim_16}, the voting rule $\nu$ is self-stable if and only if players' utility function is parametrized by $r=1$. However, since $\nu\in \mathcal{V}\backslash\mathcal{V}_v$, it follows from \Cref{t.min} that the voting rule $\nu$ is not \hyperref[sta]{minimal stable} for any preference profile satisfying \cref{ano,null,sw,dom}.
\end{exampleb}

\subsection{Alternative stability notions}\label{sub:alt}

In this section, I provide two alternative stability definitions and explore their consequences. First, I study how results change if the definition of \hyperref[sta]{minimal stability} is modified to require the absence of Pareto dominating voting rules for some (\emph{not necessarily minimal}) winning coalition.

\renewcommand{\thedefinition}{\arabic{definition}'}
\setcounter{definition}{2}

\begin{definition}[Winning stability]\label{sta'}
A voting rule $\nu$ is \emph{winning stable in $R$} if and only if there exists no voting rule that Pareto dominates it for any of its winning coalitions. Formally, if and only if
\begin{gather}\tag{ws}\label{eq:sta2}
    (\forall\nu'\in \mathcal{V})[(\nexists T\in\mathcal{W}(\nu))(\nu' \succcurlyeq_T\nu)]
\end{gather}   
\end{definition}

\renewcommand{\thedefinition}{\arabic{definition}}%

Let $\mathcal{V}'_s(R)=\{\nu\in \mathcal{V}\mid\nu\text{ is \hyperref[sta']{winning stable in $R$}}\}$.

\begin{corollary}\label{colwin}
    Let players' preference profile satisfy the \hyperref[dom]{non-dominance}, \hyperref[ano]{anonymity}, \hyperref[null]{null player} and \hyperref[sw]{swing player} axioms. Then,
\begin{enumerate}
    \item any \hyperref[oli]{oligarchic} voting rule is \hyperref[sta']{winning stable},
    \item any \hyperref[sta']{winning stable} voting rule is \hyperref[oli]{oligarchic}, or \hyperref[vet]{veto} without null players.
\end{enumerate}    
    Formally,
\begin{gather}
    (\forall R\in \mathcal{R}^*)[\mathcal{V}_o\subseteq \mathcal{V}'_s(R)\subseteq \mathcal{V}_o\cup \{\nu\in \mathcal{V}_v\mid N(\nu)=\emptyset\}]
\end{gather}
\end{corollary}

\begin{cproof}
    Consider any preference profile $R\in\mathcal{R}^*$. Then, there are two statements to show:
\begin{enumerate}    
    \item $\mathcal{V}_o\subseteq \mathcal{V}'_s(R)$,
    \item$\mathcal{V}'_s(R)\subseteq \mathcal{V}_o\cup\{\nu\in \mathcal{V}_v\mid N(\nu)=\emptyset\}$.
\end{enumerate}

\begin{statement}
    $\mathcal{V}_o\subseteq \mathcal{V}'_s(R)$. 
\end{statement}

    The proof is direct. Consider any \hyperref[oli]{oligarchic} voting rule $\nu\in \mathcal{V}_o$. By the \hyperref[sw]{swing player} axiom, there exists some voting rule $\nu'\in \mathcal{V}$ satisfying $\nu\neg I_i\nu'$ for some player $i\in N$. Fix any such voting rule $\nu'\in \mathcal{V}$. By the \hyperref[dom]{non-dominance} axiom, there exist two players $i,j\in N$ such that $\nu P_i\nu'$ and $\nu' P_j\nu$. By the \hyperref[null]{null player} axiom, $\nu'I_i\nu$ for all players $i\in N(\nu')\backslash O(\nu)$. By the \hyperref[sw]{swing player} axiom, $\nu' P_i\nu$ for all players $i\in S(\nu')\backslash O(\nu)$. Since $N(\nu')\cup S(\nu')=N$, it follows that $\nu'R_i\nu$ for all non-oligarchic players $i\in N\backslash O(\nu)$. Hence, $\nu P_i\nu'$ for some oligarchic player $i\in O(\nu)$. Since $O(\nu)\subseteq T$ for all winning coalitions $T\in\mathcal{W}(\nu)$, it follows that \cref{eq:sta2} is satisfied. Therefore, $\nu\in \mathcal{V}'_s(R)$. Thus, $\mathcal{V}_o\subseteq \mathcal{V}'_s(R)$.

\begin{statement}
    $\mathcal{V}'_s(R)\subseteq \mathcal{V}_o\cup \{\nu\in \mathcal{V}_v\mid N(\nu)=\emptyset\}$. 
\end{statement}
    
    Since $\mathcal{V}\backslash(\mathcal{V}_o\cup\{\nu\in \mathcal{V}_v\mid N(\nu)=\emptyset\})=(\mathcal{V}\backslash \mathcal{V}_v)\cup\{\nu\in \mathcal{V}_v\backslash \mathcal{V}_o\mid N(\nu)\neq\emptyset\}$, there are two claims to show:
\begin{enumerate}[label=2.\arabic*.] 
    \item $\mathcal{V}\backslash \mathcal{V}_v\subseteq \mathcal{V}\backslash \mathcal{V}'_s(R)$,
    \item $\{\nu\in \mathcal{V}_v\backslash \mathcal{V}_o\mid N(\nu)\neq\emptyset\}\subseteq \mathcal{V}\backslash \mathcal{V}'_s(R)$.
\end{enumerate}

\begin{claim}\label{cor:c21}
    $\mathcal{V}\backslash \mathcal{V}_v\subseteq \mathcal{V}\backslash \mathcal{V}'_s(R)$.
\end{claim}
    
    The proof is direct. Consider any non-\hyperref[vet]{veto} voting rule $\nu\in \mathcal{V}\backslash \mathcal{V}_v$. By \Cref{t.min}, $\nu\in\mathcal{V}\backslash \mathcal{V}_s(R)$. Since $\mathcal{M}(\nu)\subseteq\mathcal{W}(\nu)$, it follows that \cref{eq:sta2} is not satisfied. Thus, $\nu\in\mathcal{V}\backslash \mathcal{V}'_s(R)$. Hence, $\mathcal{V}\backslash \mathcal{V}_v\subseteq \mathcal{V}\backslash \mathcal{V}'_s(R)$.

\begin{claim}\label{cor:c22}
    $\{\nu\in \mathcal{V}_v\backslash \mathcal{V}_o\mid N(\nu)\neq\emptyset\}\subseteq \mathcal{V}\backslash \mathcal{V}'_s(R)$.
\end{claim}
    
    The proof is direct. Consider any non-\hyperref[oli]{oligarchic} \hyperref[vet]{veto} voting rule $\nu\in \mathcal{V}_v\backslash \mathcal{V}_o$ satisfying $N(\nu)\neq\emptyset$ and any minimal winning coalition $T\in\mathcal{M}(\nu)$. Since $O(\nu)=\emptyset$ and $N(\nu)\neq\emptyset$, there exists some permutation $\pi\in \Pi$ satisfying $\pi(i)=j\in N(\nu)$, $\pi(j)=i\in S(\nu)\backslash T$ and $\pi(k)=k$ for all players $k\in N\backslash\{i,j\}$. Fix any such permutation $\pi\in\Pi$. By the \hyperref[ano]{anonymity} axiom, $\nu I_k\pi\nu$ for all players $k\in T$. By the \hyperref[sw]{swing player} axiom, $\pi\nu P_j\nu$ and $\nu P_i\pi\nu$. Let $T'=T\cup\{j\}$. Since player $i\in N\backslash T'$, it follows that $\pi\nu\succcurlyeq_{T'}\nu$. Since $T'\in\mathcal{W}(\nu)$, it follows that \cref{eq:sta2} is not satisfied. Therefore, $\nu\in\mathcal{V}\backslash \mathcal{V}'_s(R)$. Hence, $\{\nu\in \mathcal{V}_v\backslash \mathcal{V}_o\mid N(\nu)\neq\emptyset\}\subseteq \mathcal{V}\backslash \mathcal{V}'_s(R)$.
\end{cproof}

I now study what happens if the definition of \hyperref[sta]{minimal stability} is modified to require the absence of strongly Pareto dominating voting rules for some minimal winning coalition. To do so, let $\nu\succ_T\nu'$ if and only if $\nu$ \emph{strongly Pareto dominates $\nu'$ for coalition $T$ at preference profile $R$}: formally,
\begin{gather}
    (\nu\succ_T\nu')\iff[(\forall i\in T)(\nu P_i\nu')]
\end{gather}
Further, given any coalition $T\subseteq N$, let $\Pi_T=\{\pi_T:T\to T\mid\pi_T\text{ is bijective}\}$ be the \emph{set of all permutations of coalition $T$}. Given any voting rule $\nu\in\mathcal{V}$, a coalition $T\subseteq N$ is \emph{symmetric at $\nu$} if and only if for all permutations $\pi_T\in\Pi_T$, there exists some permutation $\pi_{N\backslash T}\in\Pi_{N\backslash T}$ such that the permutation $\pi=\pi_T\cup\pi_{N\backslash T}$ satisfies $\pi\nu=\nu$.

Let $\mathcal{S}(\nu)=\{T\subseteq N\mid T\text{ is symmetric at }\nu\}$.

\renewcommand{\thedefinition}{\arabic{definition}''}
\setcounter{definition}{2}

\begin{definition}[Weak minimal stability]\label{sta''}
A voting rule $\nu$ is \emph{weak minimal stable in $R$} if and only if there exists no voting rule that strongly Pareto dominates it for any of its minimal winning coalitions. Formally, if and only if
\begin{gather}\tag{wms}\label{eq:sta3}
    (\forall\nu'\in \mathcal{V})[(\nexists T\in\mathcal{M}(\nu))(\nu' \succ_T\nu)]
\end{gather}   
\end{definition}

\renewcommand{\thedefinition}{\arabic{definition}}%
Let $\mathcal{V}''_s(R)=\{\nu\in \mathcal{V}\mid\nu\text{ is \hyperref[sta'']{weak minimal stable in $R$}}\}$.

\begin{corollary}\label{colweak}
    Let players' preference profile satisfy the \hyperref[dom]{non-dominance}, \hyperref[ano]{anonymity}, \hyperref[null]{null player} and \hyperref[sw]{swing player} axioms. Then,
\begin{enumerate}
    \item any \hyperref[oli]{oligarchic} voting rule is \hyperref[sta'']{weak minimal stable},
    \item any \hyperref[sta'']{weak minimal stable} voting rule is \hyperref[oli]{oligarchic}, or has no symmetric minimal winning coalition.
\end{enumerate}    
    Formally,
\begin{gather}
    (\forall R\in \mathcal{R}^*)[\mathcal{V}_o\subseteq \mathcal{V}''_s(R)\subseteq\mathcal{V}_o\cup\{\nu\in\mathcal{V}\mid \mathcal{S}(\nu)\cap\mathcal{M}(\nu)=\emptyset\}]
\end{gather}
\end{corollary}

\begin{cproof}
    Consider any preference profile $R\in\mathcal{R}^*$. Then, there are two statements to show: 
\begin{enumerate}
    \item $\mathcal{V}_o\subseteq\mathcal{V}_s''(R)$,
    \item $\mathcal{V}''_s(R)\subseteq\mathcal{V}_o\cup\{\nu\in\mathcal{V}\mid \mathcal{S}(\nu)\cap\mathcal{M}(\nu)=\emptyset\}$.
\end{enumerate}    

\begin{statement}
    $\mathcal{V}_o\subseteq\mathcal{V}_s''(R)$.
\end{statement}

    The proof is direct. Consider any \hyperref[oli]{oligarchic} voting rule $\nu\in\mathcal{V}_o$. By \Cref{t.min}, $\nu\in\mathcal{V}_s(R)$. If \cref{eq:sta} is satisfied, so is \cref{eq:sta3}. Hence, $\mathcal{V}_s(R)\subseteq\mathcal{V}''_s(R)$. Thus, $\nu\in\mathcal{V}_s''(R)$. Therefore, $\mathcal{V}_o\subseteq\mathcal{V}''_s(R)$. 

\begin{statement}
    $\mathcal{V}''_s(R)\subseteq\mathcal{V}_o\cup\{\nu\in\mathcal{V}\mid \mathcal{S}(\nu)\cap\mathcal{M}(\nu)=\emptyset\}$.
\end{statement}
    
    The proof is by contraposition. Consider any non-\hyperref[oli]{oligarchic} voting rule $\nu\in\mathcal{V}\backslash\mathcal{V}_o$ satisfying $\mathcal{S}(\nu)\cap\mathcal{M}(\nu)\neq\emptyset$. Fix any symmetric minimal winning coalition $T\in\mathcal{S}(\nu)\cap\mathcal{M}(\nu)$ and the unique \hyperref[oli]{oligarchic} voting rule $\nu'\in\mathcal{V}_o$ satisfying $\mathcal{M}(\nu')=\{T\}$. Since $\nu\in\mathcal{V}\backslash\mathcal{V}_o$, it follows that $N(\nu)\subsetneq N(\nu')=T^c$. Hence, the \hyperref[null]{null player} axiom implies $\nu'I_i\nu$ for all null players $i\in N(\nu)$, whereas the \hyperref[sw]{swing player} axiom implies $\nu P_i\nu'$ for all players $i\in S(\nu)\cap N(\nu')$. Thus, $\nu\succcurlyeq_{T^c}\nu'$. Then, by the \hyperref[dom]{non-dominance} axiom, $\nu' P_i\nu$ for some player $i\in T$. Since $T\in\mathcal{S}(\nu)$, the \hyperref[ano]{anonymity} axiom implies $\nu'P_i\nu$ for all players $i\in T$. Hence, $\nu'\succ_T\nu$. Since $T\in\mathcal{M}(\nu)$, it follows that \cref{eq:sta3} is not satisfied. Thus, $\nu\in\mathcal{V}\backslash\mathcal{V}''_s(R)$. Therefore, $\mathcal{V}''_s(R)\subseteq\mathcal{V}_o\cup\{\nu\in\mathcal{V}\mid \mathcal{S}(\nu)\cap\mathcal{M}(\nu)=\emptyset\}$.
\end{cproof}

\section{Minimal self-stable constitutions}\label{sec:self}

\Cref{sec:self} characterizes \hyperref[ssta]{minimal self-stable} constitutions. In order to do so, let the $2$-tuple $c=(\nu_o,\nu_e)\in \mathcal{V}^2$ be a \emph{constitution}, where $\nu_o$ is the \emph{ordinary voting rule} used to decide over quotidian matters, and $\nu_e$ is the \emph{extraordinary voting rule} used to amend either of its voting rules, \emph{including itself}.

Let $\mathcal{C}=\{c\mid c\text{ is a constitution}\}$.

\begin{definition}[Minimal self-stability]\label{ssta}
    A constitution $c=(\nu_o,\nu_e)$ is \emph{minimal self-stable in $R$} if and only if there exists no voting rule that Pareto dominates neither its ordinary nor its extraordinary voting rule for any minimal winning coalition of its extraordinary voting rule. Formally, if and only if
\begin{gather}\tag{mss}\label{eq:con}
    (\forall\nu'\in \mathcal{V})(\forall \nu\in\{\nu_o,\nu_e\})[(\nexists T\in\mathcal{M}(\nu_e))(\nu'\succcurlyeq_T\nu)]
\end{gather}
\end{definition}

Let $\mathcal{C}_s(R)=\{c\in \mathcal{C}\mid c\text{ is \hyperref[ssta]{minimal self-stable in $R$}}\}$.

The definition of \hyperref[ssta]{minimal self-stability} is a straightforward extension of \hyperref[sta]{minimal stability} to pairs of voting rules in which constitutional amendments are governed by the extraordinary voting rule $\nu_e$. Apart from the use of standard---rather than strong---Pareto domination, the main novelty of this \hyperref[ssta]{minimal self-stability} notion is the requirement that the extraordinary voting rule $\nu_e$ be \hyperref[sta]{minimal stable} for a constitution to be \hyperref[ssta]{minimal self-stable}. This requirement stems from the fact that, in most constitutions around the world, the extraordinary voting rule $\nu_e$ can be used to amend itself. Without this requirement, the extraordinary voting rule $\nu_e$ could be amended in such a way that a coalition that desires to change the ordinary voting rule $\nu_o$ but could not do so before the amendment of $\nu_e$ can do so after the amendment of $\nu_e$. In light of this observation, it may be problematic to drop the requirement that the extraordinary voting rule $\nu_e$ be \hyperref[sta]{minimal stable} for a constitution to be considered \hyperref[ssta]{minimal self-stable}.

\begin{theorem}\label{t.con}
    Let players' preference profile satisfy the \hyperref[dom]{non-dominance}, \hyperref[ano]{anonymity}, \hyperref[null]{null player} and \hyperref[sw]{swing player} axioms. Then, a constitution is
\begin{enumerate}
    \item \hyperref[ssta]{minimal self-stable} if its ordinary swing player set is a subset of its extraordinary oligarchic player set (\cref{eq:consta}),
    \item not \hyperref[ssta]{minimal self-stable} if its extraordinary veto player set is empty, or is neither a superset nor a subset of its ordinary swing player set (\cref{eq:conunsta}).
\end{enumerate}
    Formally,
\begin{gather}\label{eq:consta}
    (\forall R\in \mathcal{R}^*)[\{c\in \mathcal{C}\mid S(\nu_o)\subseteq O(\nu_e)\}\subseteq \mathcal{C}_s(R)]\\\label{eq:conunsta}
    (\forall R\in \mathcal{R}^*)[\{c\in \mathcal{C}\mid (V(\nu_e)=\emptyset)\vee(S(\nu_o)\not\subseteq V(\nu_e)\not\subseteq S(\nu_o))\}\subseteq \mathcal{C}\backslash \mathcal{C}_s(R)]
\end{gather}
\end{theorem}

\begin{proof}
    Consider any preference profile $R\in \mathcal{R}^*$. Then, there are two statements to show:
\begin{enumerate}    
    \item $\{c\in \mathcal{C}\mid S(\nu_o)\subseteq O(\nu_e)\}\subseteq \mathcal{C}_s(R)$,
    \item $\{c\in \mathcal{C}\mid (V(\nu_e)=\emptyset)\vee(S(\nu_o)\not\subseteq V(\nu_e)\not\subseteq S(\nu_o))\}\subseteq \mathcal{C}\backslash \mathcal{C}_s(R)$.
\end{enumerate}

\begin{statement}
    $\{c\in \mathcal{C}\mid S(\nu_o)\subseteq O(\nu_e)\}\subseteq \mathcal{C}_s(R)$.
\end{statement}
    
    Consider any constitution $c=(\nu_o,\nu_e)$ satisfying $S(\nu_o)\subseteq O(\nu_e)$. Then, there are two claims to show:
\begin{enumerate}[label=1.\arabic*.]
    \item $(\forall\nu'\in \mathcal{V})[(\nexists T\in\mathcal{M}(\nu_e))(\nu'\succcurlyeq_T\nu_o)]$,
    \item $(\forall\nu'\in \mathcal{V})[(\nexists T\in\mathcal{M}(\nu_e))(\nu'\succcurlyeq_T\nu_e)]$.
\end{enumerate}

\begin{claim}\label{t.con:c11}
    $(\forall\nu'\in \mathcal{V})[(\nexists T\in\mathcal{M}(\nu_e))(\nu'\succcurlyeq_T\nu_o)]$.
\end{claim}

    The proof is direct. By the \hyperref[sw]{swing player} axiom, there exists some voting rule $\nu'\in \mathcal{V}$ satisfying $\nu_o\neg I_i\nu'$ for some player $i\in N$. Fix any such voting rule $\nu'\in \mathcal{V}$. By the \hyperref[dom]{non-dominance} axiom, there exist two players $i,j\in N$ such that $\nu_o P_i\nu'$ and $\nu' P_j\nu_o$. By the \hyperref[null]{null player} axiom, $\nu' I_i\nu_o$ for all players $i\in N(\nu_o)\cap N(\nu')$. By the \hyperref[sw]{swing player} axiom, $\nu' P_i\nu_o$ for all players $i\in N(\nu_o)\cap S(\nu')$. Since $N(\nu')\cup S(\nu')=N$, it follows that $\nu'R_i\nu_o$ for all ordinary null players $i\in N(\nu_o)$. Therefore, $\nu_o P_i\nu'$ for some ordinary swing player $i\in S(\nu_o)$. Since $S(\nu_o)\subseteq O(\nu_e)$, it follows that $\nu_o P_i\nu'$ for some extraordinary oligarchic player $i\in O(\nu_e)$. Since $\mathcal{M}(\nu_e)=\{O(\nu_e)\}$, there exists no extraordinary minimal winning coalition $T\in\mathcal{M}(\nu_e)$ satisfying $\nu'\succcurlyeq_T\nu_o$. 
    
\begin{claim}\label{t.con:c12}
    $(\forall\nu'\in \mathcal{V})[(\nexists T\in\mathcal{M}(\nu_e))(\nu'\succcurlyeq_T\nu_e)]$.
\end{claim}

    The proof is direct. Since $\nu_e\in \mathcal{V}_o$, \Cref{t.min} implies that there exists no voting rule $\nu'\in \mathcal{V}$ satisfying $\nu'\succcurlyeq_T\nu_e$ for some extraordinary minimal winning coalition $T\in\mathcal{M}(\nu_e)$. 

\begin{statement}
    $\{c\in \mathcal{C}\mid (V(\nu_e)=\emptyset)\vee(S(\nu_o)\not\subseteq V(\nu_e)\not\subseteq S(\nu_o))\}\subseteq \mathcal{C}\backslash \mathcal{C}_s(R)$. 
\end{statement}

    There are two claims to show:
\begin{enumerate}[label=2.\arabic*.]
    \item $\{c\in \mathcal{C}\mid V(\nu_e)=\emptyset\}\subseteq \mathcal{C}\backslash \mathcal{C}_s(R)$,
    \item $\{c\in \mathcal{C}\mid S(\nu_o)\not\subseteq V(\nu_e)\not\subseteq S(\nu_o)\}\subseteq \mathcal{C}\backslash \mathcal{C}_s(R)$.
\end{enumerate}

\begin{claim}\label{t.con:c21}
    $\{c\in \mathcal{C}\mid V(\nu_e)=\emptyset\}\subseteq \mathcal{C}\backslash \mathcal{C}_s(R)$.
\end{claim}
    
    The proof is direct. Consider any constitution $c\in\mathcal{C}$ satisfying $V(\nu_e)=\emptyset$. Then, since $\nu_e\in\mathcal{V}\backslash \mathcal{V}_v$, \Cref{t.min} implies that there exists some voting rule $\nu'\in \mathcal{V}$ satisfying $\nu'\succcurlyeq_T\nu_e$ for some extraordinary minimal winning coalition $T\in\mathcal{M}(\nu_e)$. Therefore, \cref{eq:con} is not satisfied. Hence, $c\in\mathcal{C}\backslash \mathcal{C}_s(R)$. Thus, $\{c\in\mathcal{C}\mid V(\nu_e)=\emptyset\}\subseteq\mathcal{C}\backslash\mathcal{C}_s(R)$. 

\begin{claim}\label{t.con:c22}
    $\{c\in \mathcal{C}\mid S(\nu_o)\not\subseteq V(\nu_e)\not\subseteq S(\nu_o)\}\subseteq \mathcal{C}\backslash \mathcal{C}_s(R)$.
\end{claim}
    
    The proof is direct. Consider any constitution $c=(\nu_o,\nu_e)$ satisfying $S(\nu_o)\not\subseteq V(\nu_e)\not\subseteq S(\nu_o)$. Then, $\nu_e\in V_v$. Since $S(\nu_o)\not\subseteq V(\nu_e)$, it follows that $V(\nu_e)^c\cap S(\nu_o)\neq\emptyset$; and since $V(\nu_e)\not\subseteq S(\nu_o)$, it follows that $V(\nu_e)\cap N(\nu_o)\neq\emptyset$. Fix any extraordinary veto and ordinary null player $i\in V(\nu_e)\cap N(\nu_o)$ and any extraordinary non-veto and ordinary swing player $j\in V(\nu_e)^c\cap S(\nu_o)$. Then, consider the voting rule $\pi\nu_o\in \mathcal{V}$, where $\pi(i)=j$, $\pi(j)=i$ and $\pi(k)=k$ for all players $k\in N\backslash\{i,j\}$. By the \hyperref[sw]{swing player} axiom, $\pi\nu_o P_i\nu_o$ and $\nu_o P_j\pi\nu_o$. By the \hyperref[ano]{anonymity} axiom, $\pi\nu_o I_k\nu_o$ for all players $k\in N\backslash \{i,j\}$. Since player $j\in N\backslash V(\nu_e)$, there exists some extraordinary minimal winning coalition $T\in\mathcal{M}(\nu_e)$ for which player $j\in N\backslash T$. Since player $i\in V(\nu_e)$, it follows that $\pi\nu_o\succcurlyeq_T\nu_o$ for all extraordinary minimal winning coalitions $T\in\mathcal{M}(\nu_e)$ such that player $j\in N\backslash T$. Then, \cref{eq:con} is not satisfied. Hence, $c\in\mathcal{C}\backslash \mathcal{C}_s(R)$. Thus, $\{c\in \mathcal{C}\mid S(\nu_o)\not\subseteq V(\nu_e)\not\subseteq S(\nu_o)\}\subseteq \mathcal{C}\backslash \mathcal{C}_s(R)$.
\end{proof}

Finally, in \Cref{ex3,ex4}, I illustrate how the results in \Cref{t.con} compare with those of \textcite[Propositions 2 to 5, p. 385]{azrielikim_16}.

\begin{exampleb}\label{ex3}
    Consider a society with three players $N=\{1,2,3\}$ and the constitution $c\in\mathcal{C}$ given by the sets of minimal winning coalitions $\mathcal{M}(\nu_e)=\mathcal{M}(\nu_o)=\{\{1\}\}$. By \textcite[Propositions 2 to 5, p. 385]{azrielikim_16}, the constitution $c$ might (or might not) be self-stable in their setup. However, since $O(\nu_o)=O(\nu_e)=\{1\}$, it follows from \Cref{t.con} that the constitution $c$ is \hyperref[ssta]{minimal self-stable} for all preference profiles satisfying \cref{ano,null,sw,dom}. 
\end{exampleb}

\begin{exampleb}\label{ex4}
    Consider a society with three players $N=\{1,2,3\}$ and any constitution $c\in\mathcal{C}$ given by the sets of minimal winning coalitions $\mathcal{M}(\nu_e)=\mathcal{M}(\nu_o)=\{\{1,2\},\{1,3\},\{2,3\}\}$. By \textcite[Propositions 2 to 5, p. 385]{azrielikim_16}, the constitution $c$ might (or might not) be self-stable in their setup. However, since $\nu_e\in\mathcal{V}\backslash\mathcal{V}_v$, it follows from \Cref{t.con} that the constitution $c$ is not \hyperref[ssta]{minimal self-stable} for any preference profile satisfying \cref{ano,null,sw,dom}. 
\end{exampleb}

\section{Conclusion}\label{sec:con}

In general, the \hyperref[ssta]{minimal self-stability} of most constitutions is a complex issue to address. On the one hand, some constitutions belong to the category for which this paper does not have a definitive answer (\Cref{exun}). On the other hand, the \hyperref[ssta]{minimal self-stability} of many constitutions depends on who the real players are (\Cref{exchina,exusa}). If individual legislators are assumed to be the real players, most constitutions around the world are not \hyperref[ssta]{minimal self-stable} when their preference profile satisfies \cref{ano,sw,null,dom}. However, if political parties are assumed to be the real players, many constitutions around the world are \hyperref[ssta]{minimal self-stable} when their preference profile satisfies \cref{ano,sw,null,dom}.

One shall expect any non-\hyperref[ssta]{minimal self-stable} constitution to naturally evolve until becoming \hyperref[ssta]{minimal self-stable}. Clearly, this transformation can happen through successive constitutional amendments; but it can also happen through the emergence of political parties that---by coordinating their respective members---end up acting as oligarchic players. This latter view is supported by the fact that party discipline generally forces all parliamentary members of the same party to vote together for the same option. In other words, party discipline transfers the ability to choose from the parliamentary members to their respective parties. And since a player is only so if it is able to decide for itself, it is entirely reasonable to think of political parties as the true players. Therefore, the reader can understand the results of this paper as suggesting that most constitutions around the world are not \hyperref[ssta]{minimal self-stable}, or as suggesting that political parties have evolved to ensure the \hyperref[ssta]{minimal self-stability} of otherwise not \hyperref[ssta]{minimal self-stable} constitutions.

\begin{example}[Charter of the United Nations \citeyearpar{uncharter_45}]\label{exun}
    The \textcite{uncharter_45} sets two different voting rules: simple majority of the UN General Assembly for most ordinary issues,\footnote{Some issues require a two-thirds majority, but this does not alter the conclusion of \Cref{exun}.} and a more stringent voting rule for amendments. According to its Article 108, Chapter XVIII; amendments are passed if and only if they are approved by two thirds of the UN General Assembly, including the five UN Security Council permanent members. At the time of writing this paper, a coalition is winning in its extraordinary voting rule if and only if it contains at least $(2/3)193\approx 129$ member states, including France, China, Russia, the UK and the US. Therefore, these five states are extraordinary non-oligarchic veto players, whereas all member states are ordinary swing players. Thus, by \Cref{t.con}, the \textcite{uncharter_45} may (or may not) be \hyperref[ssta]{minimal self-stable}.
\end{example}

\begin{example}[Constitution of the People's Republic of China \citeyearpar{chinaconstitution_82}]\label{exchina}
    The \textcite{chinaconstitution_82} sets two different voting rules: simple majority of the National People's Congress (NPC) for ordinary issues, and a qualified majority for amendments.\footnote{Since the NPC only meets yearly, a lot of ordinary legislation is passed by its 175-member Standing Committee, whose partisan composition mirrors that of the NPC.} According to its Article 64, Chapter 3; its reform requires the approval by two thirds of the NPC. The 13th NPC had $2,980$ legislators.\footnote{At the time of writing this paper, the partisan composition of the 14th NPC has not been publicly released.} Under the assumption that political parties---rather than individual legislators---are the real players, a coalition was winning in both voting rules if and only if it included the Chinese Communist Party, as it controlled $2,119>(2/3)2,980\approx 1,987$ legislators. Hence, the Chinese Communist Party was the unique oligarchic player in both voting rules. Thus, by \Cref{t.con}, the \textcite{chinaconstitution_82} was \hyperref[ssta]{minimal self-stable}.
\end{example}

\begin{example}[Constitution of the United States of America \citeyearpar{usconstitution_87}]\label{exusa}
    The \textcite{usconstitution_87} sets two different voting rules: simple majority of both chambers of the US Congress for ordinary issues,\footnote{The US President can veto legislation, but the Congress can override this veto by approving said legislation with a two-thirds majority in each of its chambers. However, this does not alter the conclusion of \Cref{exusa}.} and a more stringent voting rule for amendments. According to its Article V, its reform requires the approval by two-thirds of both chambers of the US Congress and by three-fourths of US state legislatures.\footnote{Article V of the \textcite{usconstitution_87} allows the US Congress to let states ratify amendments by convention, but this process has only been used once for the ratification of the 21st Amendment in 1933, and is thus omitted from \Cref{exusa} for simplicity.} At the time of writing this paper, the Senate has $100$ senators, the House has $435$ representatives, and the US has $50$ states. Under the assumption that political parties---rather than individual legislators---are the real players, a coalition is winning in its extraordinary voting rule if and only if it contains both the Democratic and the Republican parties, as the former controls $45$ senators, $213$ representatives and $18$ state legislatures; and the latter controls $53$ senators, $218$ representatives and $28$ state legislatures.\footnote{\Cref{exusa} assumes that ratification by each state legislature is by simple majority, but larger majorities may be required in some states. However, this simplification does not alter the conclusion of \Cref{exusa}.} Hence, the Democratic and Republican parties form a two-player \hyperref[oli]{oligarchy} in the extraordinary voting rule, whereas the Republican Party is the unique oligarchic player of the ordinary voting rule. Thus, by \Cref{t.con}, the \textcite{usconstitution_87} is \hyperref[ssta]{minimal self-stable}.
\end{example}

\acknowledgments{I am profoundly grateful to my two PhD supervisors, Ben McQuillin and Mich Tvede, for countless discussions that were fundamental to the development of this paper. I am also indebted for their help to Roberto Serrano, the Advisory Editor, several anonymous referees, Ferran Hermida Rivera, Jos\'{e} Mar\'{i}a Alonso-Meijide, Vicen\c c Esteve Guasch, Josep Freixas, Faruk G\"{u}l, B\aa rd Harstad, Toygar T. Kerman, L\'{a}szl\'{o} \'{A}. K\'{o}czy, Marc Claveria-Mayol, Michele Lombardi, Andrew Mackenzie, Andrea Marietta Leina, Bruno Strulovici, Robert Sugden, QSMS members, and participants in the University of East Anglia School of Economics Seminars, the Corvinus University of Budapest Game Theory Seminars, the University of Liverpool Online Economic Theory Seminars, the University of Vienna Department of Economics Research Privatissimum, the 26th \& 27th Coalition Theory Network Workshops, the 12th Lisbon Meetings in Game Theory \& Applications, the 13th Conference on Economic Design, the 18th European Meeting on Game Theory, and the 2024 Conference on Mechanism \& Institution Design. All errors are only mine. A previous draft of this paper was the central chapter of my PhD Thesis. I acknowledge financial support of a PhD scholarship awarded by the Faculty of Social Sciences of the University of East Anglia. I declare no conflict of interest.}

\conflictofinterest{I declare no conflict of interest.}

\dataavailability{No data was used to write this paper.}

\funding{None.}

\printbibliography[]
\end{document}